\documentclass[sn-basic,Numbered]{sn-jnl}% Basic 

\usepackage{graphicx}%
\usepackage{multirow}%
\usepackage{amsmath,amssymb,amsfonts}%
\usepackage{amsthm}%
\usepackage{multirow}
\usepackage{mathrsfs}%
\usepackage[title]{appendix}%
\usepackage{hhline}
\usepackage{textcomp}%
\usepackage{manyfoot}%
\usepackage{booktabs}%
\usepackage{algorithm}%
\usepackage{algorithmicx}%
\usepackage{algpseudocode}%
\usepackage{listings}%
\usepackage{siunitx}%
\usepackage[table,xcdraw]{xcolor}
\usepackage{colortbl}
\usepackage{caption}
\usepackage{soul}
\usepackage{ulem}
\usepackage[version=3]{mhchem}

\begin{document}

\title[]{Silicon photonic modulator circuit with programmable intensity and phase modulation response
}

\author*[1,2]{\fnm{Hong} \sur{Deng}}\email{hong.deng@ugent.be}
% \equalcont{These authors contributed equally to this work.}
\author[1,2]{\fnm{Yu} \sur{Zhang}}\email{yzhangzh.zhang@ugent.be}
% \equalcont{These authors contributed equally to this work.}
\author[1,2]{\fnm{Xiangfeng} \sur{Chen}}\email{xiangfeng.chen@ugent.be}
\author*[1,2]{\fnm{Wim} \sur{Bogaerts}}\email{wim.bogaerts@ugent.be}
\affil[1]{\orgdiv{Photonics Research Group, Department of Information Technology, },\orgname{Ghent University - imec}, \orgaddress{\city{Ghent}, \country{Belgium}}}

\affil[2]{Center for Nano- and Biophotonics (NB Photonics), Ghent University, \orgaddress{\city{Ghent}, \country{Belgium}}}

\abstract{Electro-optical modulators are essential components in optical communication systems. They encode an electrical waveform onto an optical carrier. However, their performance is often limited by inherent electro-optic processes and imperfections in existing integrated designs, which limit their adaptability to diverse applications. This paper presents a circuit-level programmable modulator design that addresses these challenges. The proposed modulator can generate  both intensity and phase modulation, optimizing performance without altering the underlying design or constraining platform limitations. We explain and demonstrate the principle with both carrier depletion-based modulators and SiGe electro-absorption modulators on a silicon photonic platform. Experiments demonstrate precise control and optimization capabilities surpassing those of traditional modulator designs, marking a significant leap forward in adaptability and performance enhancement across intensity, phase, and modulation linearity, enabled by programmable photonics. In combination with on-chip monitors, our circuit can be self-calibrating. This programmable modulator circuit concept can be applied to modulators in different platforms, incorporating  existing phase shifter designs, and act as a drop-in replacement in more complex circuits. It has the potential to be widely used in optical communication, LiDAR, microwave photonics, and other systems, meeting the increasing demands of various applications.}

\keywords{Silicon photonics, integrated photonics, electro-optical modulator}

\maketitle

\section{Introduction}\label{sec1}
Electro-optical (E/O) modulators play an important role in integrated photonic systems, converting microwave signals into the optical domain by modulating the waveform onto the phase and/or amplitude of an optical carrier wave \cite{Reed2010SiliconModulators}. As photonic chip technologies are moving beyond the field of optical fiber communication, and into  applications for optical sensing \cite{Cheng2023On-chipSensing}, LiDAR \cite{Sayyah2022FullyChip}, and microwave signal processing (so-called microwave photonic systems)\cite{Feng2024IntegratedEngine}, the requirements on the optical modulators are also changing rapidly, including, but not limited to, larger modulation bandwidths, higher modulation efficiencies,  strong linearity, and a "pure" modulation response\cite{,Thomson2016RoadmapPhotonics,Shekhar2024RoadmappingPhotonics}.

Various approaches have been proposed to achieve E/O modulators in photonic integrated circuits. For instance, in most mature silicon photonic platforms, the commonly used modulators are based on waveguides with embedded \textit{pn} junctions, which affect the propagation of light through the free carrier dispersion effect. These \textit{pn} junctions can be embedded in an interferometric circuit to form microring resonator modulators (MRM) and traveling-wave Mach-Zehnder modulators (TWMZM). MRMs normally have a larger electrical modulation bandwidth \cite{Li2020ADriver}, but can only work for specific wavelengths, and also suffer from thermal drift. TWMZMs on the other hand, can be operated over a wide wavelength range, but take up a large footprint and suffer from a relatively low electrical bandwidth. Still, with a co-designed and co-packaged modulator driver, a TWMZM can also reach a bandwidth higher than \qty{67}{GHz} \cite{Li2023AnEfficiency}. In parallel with pushing the bandwidth, various approaches to improve the linearity of these MRMs and TWMZMs have been pursued \cite{Chen2020High-LinearityStructure, Yu2020HighDistortion, Bottenfield2019SiliconLinks}, using specifically-targeted custom-designed modulators. Good linearity, where the modulator does not introduce higher harmonics and nonlinear distortion of the signal, is very desirable in analog signal processing applications such as microwave photonics. Spectrum shaping is also proposed to improve the linearity of the system, but this process is quite complicated \cite{Liu2021IntegratedEnhancement, Daulay2022UltrahighFilter, Guo2021VersatileShaper}.

Apart from these carrier dispersion-based modulators mentioned above, some silicon photonics platforms also offer SiGe electro-absorption modulators (EAM) based on the Franz-Keldysh effect \cite{Ferraro2023ImecRoadmap}. These devices typically have a much reduced footprint and larger modulation bandwidth. Alternative E/O modulators, such as InP (based on Franz-Keldysh or Pockels effect) \cite{Pintus2019CharacterizationTemperature}, thin film $\ce{LiNbO_3}$ on insulator (LNOI) \cite{Wang2018IntegratedVoltages,Vanackere2023HeterogeneousPrinting}, thin-film $\ce{BaTiO_3}$ \cite{Eltes2019APlatform} (Pockels effect), polymer modulators \cite{Taghavi2022PolymerProjections}, and silicon-organic hybrids \cite{Alloatti2014100GHzModulator}, can also been heterogeneously integrated onto silicon photonic circuits, with wafer-to-wafer, die-to-wafer or die-to-die bonding, or transfer-printing technologies \cite{Rahim2021TakingTechnologies}. These modulators also show promising modulation performance. However, such modulator designs are mainly explored for larger modulation bandwidth, driven by the needs for high-speed digital optical interconnects. Their performance metrics for analog signals, such as spurious intensity or phase modulation, modulation linearity and spurious-free dynamic range (SFDR), are usually far from optimal. None of these many E/O modulation mechanisms is perfect, and as different applications impose their own requirements, it is not straightforward to meet them within a single device. 

In this paper, we present a programmable modulator design, which takes existing E/O modulators and embeds them into a fully tunable Mach-Zehnder interferometer (MZI) circuit. By adjusting the static transmission response of this MZI, we can compensate for the imperfections in the electro-optic response of the embedded modulator, and achieve an improved modulation response curve, lower insertion loss, higher modulation efficiency, or higher SFDR, depending on the requirements of the application. Furthermore, if the embedded modulator component provides a large phase modulation, the programmable modulator circuit can be adjusted to perform as an intensity modulator, a phase modulator, or a combination of both modulation formats. Previously, we already described the static (DC) simulation results for achieving a pure phase modulation with a \textit{pn} junction based modulator \cite{Deng2019PureModulator}, and showed how this modulator can perform certain microwave photonic signal processing functions \cite{Deng2023Single-ChipSignals}. In this paper, we present a detailed experimental demonstration with \textit{pn} modulators and SiGe EAMs, supported by analytical analysis. Notably, experiments with \textit{pn} modulators achieve exceptional linearity (SFDR up to \qty{124.6}{dBHz^{2/3}}), alongside pure phase modulation (spurious intensity suppressed by \qty{20}{dB}) and high extinction (\qty{38}{dB}), across various operating conditions. In addition, this programmable modulator circuit concept can be applied to all existing platforms and E/O modulator technologies, and we show this with an additional numerical simulation and analysis of a Silicon MRM. The performance of the system can be roughly predicted from the DC response of the embedded modulator. The DC response (optical phase and amplitude transmission as function of applied voltage) can be directly characterized within our programmable modulator circuit, even in cases where the detailed design and performance specifications of the embedded modulator are not fully available up-front. With the integration of on-chip thermal phase shifters and monitor photodetectors, the proposed modulator circuit is self-calibrating.

\section{Results}\label{sec2}

\begin{figure}
    \centering
    \includegraphics[width=1\linewidth]{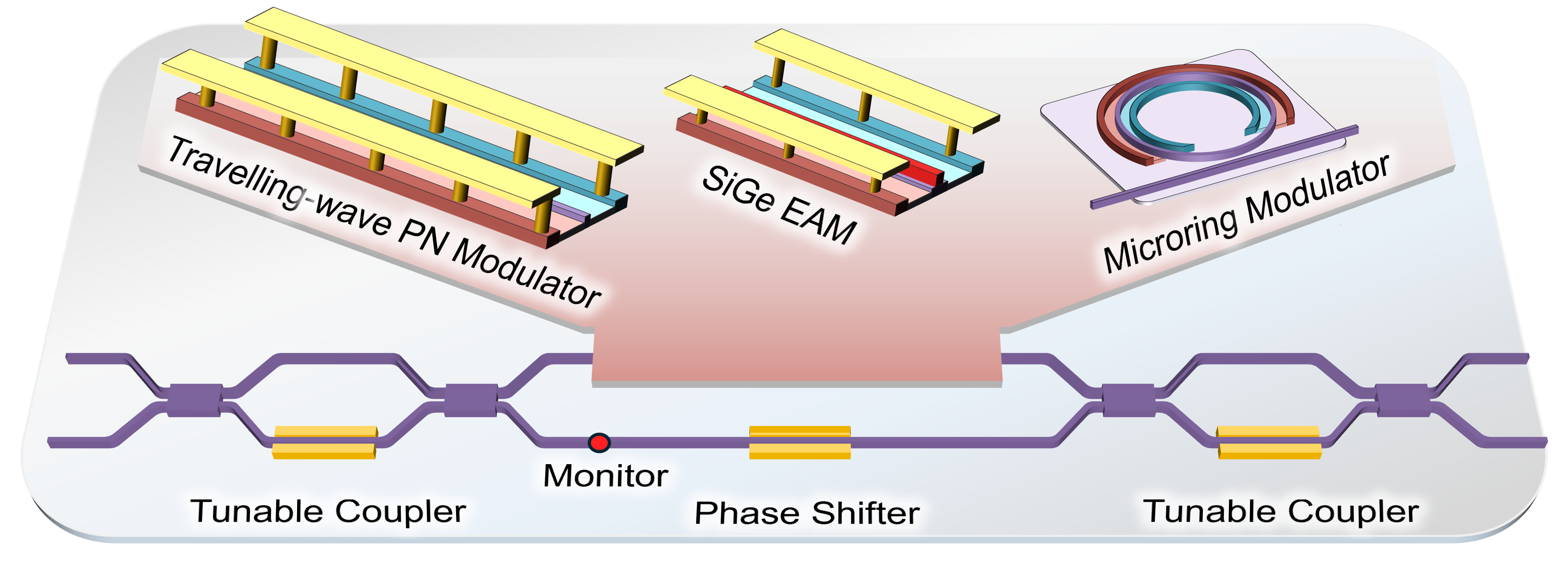}
    \caption{Schematic of the programmable modulator. The embedded high-speed modulator can be travelling wave \textit{pn} modulator, SiGe EAM modulator, microring modulator, and other modulators on silicon photonic platform or other platforms.}
    \label{f:1}
\end{figure}
The schematic of our programmable modulator is shown in Fig.~\ref{f:1}. As can be seen, the programmable modulator is an MZI circuit formed by two tunable couplers, a high-speed modulator in one arm, a static phase shifter in the other arm, and a tap monitor. The tunable coupler itself is also an MZI consisting of two 50:50 multimode interferometer (MMI) couplers and a static phase shifter. By adjusting the phase difference between the two arms of a tunable coupler, we can adjust the power splitting ratio between the two outputs. A monitor photodetector (tapping off a bit of light) is attached  to one output port to calibrate the tunable coupler. The static phase shifters on our chip are implemented by undercut heaters \cite{Ferraro2023ImecRoadmap}, but could just as well be based on liquid crystal phase shifters \cite{VanIseghem2022LowPlatform} or microelectromechanical phase shifters \cite{Edinger2021SiliconPhotonics}. The two tunable couplers, together with the static phase shifter, form an MZI structure with a fully tunable sinusoidal transmission response, which is used to compensate for the imperfections of the embedded modulator. The high-speed modulator can be a \textit{pn} junction phase modulator, an MRM, a SiGe EAM, or any other E/O modulator device. In this work, we experimentally demonstrate the performance improvements of a \textit{pn} junction modulator and a SiGe EAM, and use our circuit model to predict the performance improvements of an MRM on imec's iSiPP50G silicon photonic platform.

\subsection{PN Junction-based travelling wave modulator} 
\begin{figure}
    \centering
    \includegraphics[width = \linewidth]{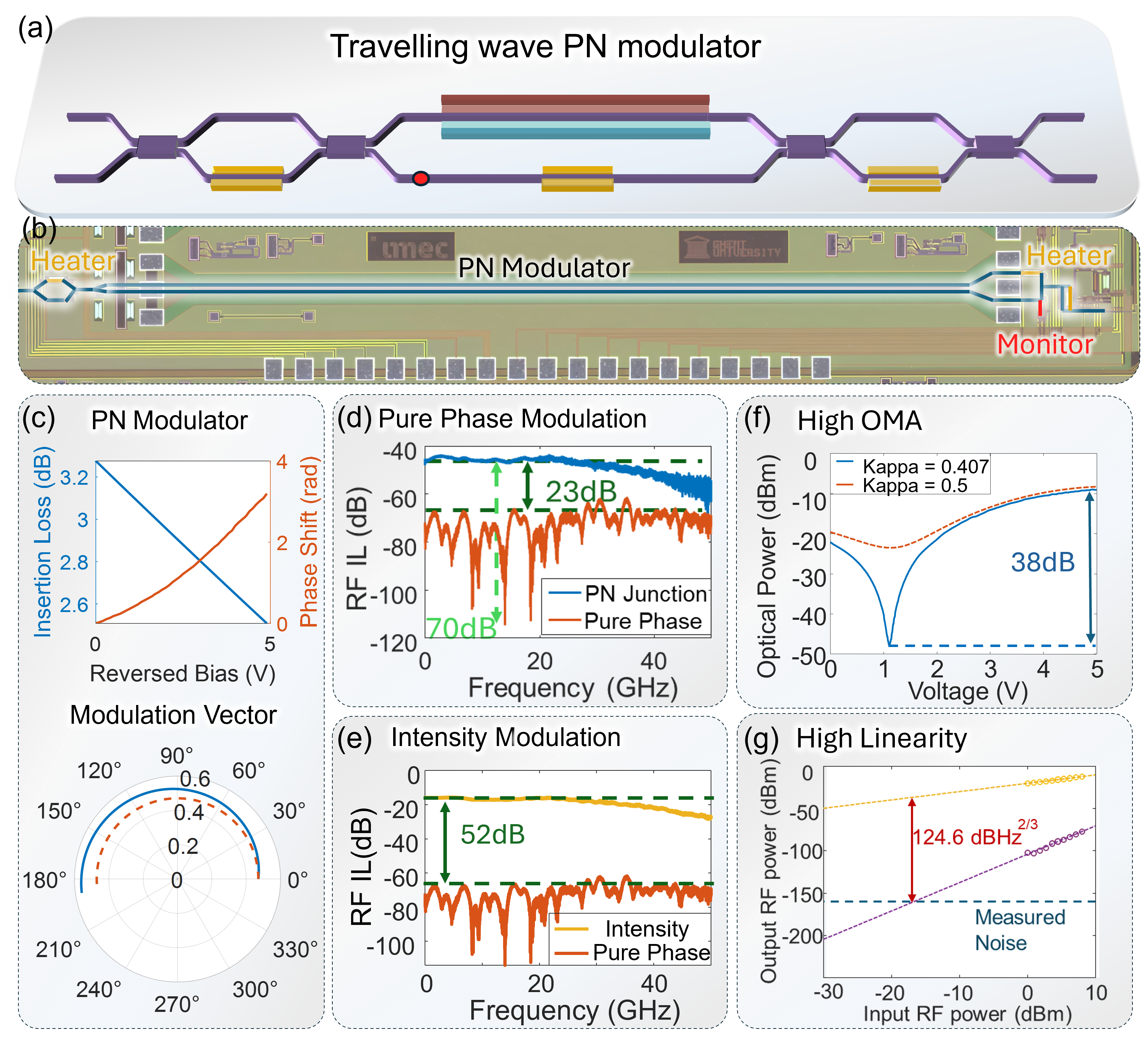}
    \caption{Programmable modulator circuit based on traveling wave \textit{pn} modulator. (a) The schematic of the modulator; (b) Fabricated demonstrator; (c) Insertion loss and the phase shift introduced by the embedded reverse biased \textit{pn} junction;(d) S21 of the optimized pure phase modulation and the original PN junction; (e)S21 of the intensity modulation and pure phase modulation; (f)DC response of the optimized high extinction ratio modulation and a normal 50:50 MZI modulator; (g) Optimized high-linearity modulation.}
    \label{f:2}
\end{figure}

The \textit{pn} junction-based traveling wave modulator is one of the most commonly used modulators in silicon photonic circuits \cite{Rahim2021TakingTechnologies}. By extracting the free carriers from a silicon waveguide core, the \textit{pn} junction can change the refractive index of the waveguide and induce a phase modulation in the guided light. However, this carrier density variance also change the insertion loss of the waveguide, resulting in a spurious intensity modulation \cite{Nedeljkovic2011Free-carrierRange}. The schematic of the \textit{pn} junction based programmable modulator circuit is shown in Fig.~\ref{f:2}(a), and a fabricated demonstrator is shown in Fig.~\ref{f:2}(b).  The intensity response and phase response of a fabricated \textit{pn} junction modulator are shown in Fig.~\ref{f:2}(c). 

\paragraph{Pure phase modulation} 
The fully programmable modulator provides a tunable sinusoidal intensity response based on the phase change of the embedded \textit{pn} junction modulator. This can be configured to have an opposite slope of the spurious intensity modulation of the \textit{pn} junction, and eventually the intensity modulation of the entire circuit is suppressed around the operating point. The experimental setup used for characterization, and the obtained measurement results are shown in Fig.~\ref{f:2}(d). With the shown setup, a phase-modulated signal will not be detected by a simple photodetector. However, due to the spurious intensity modulation of the \textit{pn} junction, an RF signal is recovered ($\kappa = 0$) with an insertion loss of \qty{45}{dB}. With a proper setting of the programmable modulator ($\kappa = 0.3$, $\phi_s = 1.5 \pi$), this transmitted RF signal (i.e. the intensity modulation) is suppressed. The generated RF signal (spurious intensity modulation) can be suppressed by \qty{23}{dB} overall, with a maximum suppression of \qty{70}{dB}.

\paragraph{High RF-gain intensity modulation}
The RF gain depends on the slope of the modulation curve, as well as the total optical loss. The highest RF gain is obtained when the programmable modulator is set to $\kappa = 0.48$, which balances both effects (more details are shown in the supplementary information). The RF transmission response is shown in Fig.~\ref{f:2}(e). From the highest RF gain intensity modulation to the lowest RF gain phase modulation, the proposed modulator shows \qty{52}{dB} suppression overall, with a maximum suppression of \qty{97}{dB} extinction ratio over the whole frequency range, which is superior to the reported \qty{64}{dB} record (only one frequency point due to optical dispersion) using spectrum shaping \cite{Daulay2022UltrahighFilter}.

\paragraph{High extinction modulation}
An intensity modulation with a high extinction ratio would obtain a high optical modulation amplitude, ensuring a high signal-to-noise ratio (SNR) in optical communication network and optical sensing system \cite{Cheng2023On-chipSensing}. For an interferometer-based modulator, the extinction ratio of the modulated signal depends on the destructive interference at the optical output. However, the extra insertion loss of the embedded \textit{pn} junction breaks the propagation loss symmetry, resulting in a low extinction ratio, if 50:50 power couplers are used. Two tunable couplers in our programmable modulator circuits can be tuned to compensate the loss imbalance of the MZI, enabling intensity modulation with high and robust extinction ratio. Experimental results are shown in Fig.~\ref{f:2}(f) As can be seen, when the programmable modulator is set at $\kappa = 0.407$ (around 60\% light is fed into the \textit{pn} junction), we can achieve a high extinction ratio (\qty{38}{dB}) in the modulator bias sweep. This result is not as good as the results shown in \cite{Cheng2023On-chipSensing} (\qty{71}{dB} in the optical DC sweep, \qty{68}{dB} in the RF tests), but in our case the programmable modulator will not be dependent on the optical carrier wavelength.

\paragraph{High-linearity intensity modulation}
As shown in Fig.~\ref{f:2}(c), the stand-alone \textit{pn} junction shows a nonlinear phase and a nonlinear spurious intensity response. In addition, the intensity modulation is simply implemented by converting the phase modulation using an MZI, which gives a sinusoidal response and is therefore intrinsically nonlinear. With our programmable modulator circuit, we can significantly linearize the intensity modulation response by combining these two sources of nonlinearity and canceling them out. The linearity of a modulator is evaluated by its spurious-free dynamic range (SFDR). The linearized result is shown in Fig.~\ref{f:2}(g). When the programmable modulator is configured with $\kappa = 0.48$ and $\phi_s = 0.31\pi$, the SFDR is recorded at \qty{124.6}{dBHz^{2/3}} around a modulation frequency of \qty{2}{GHz} (with a noise floor of \qty{-160.4}{dBm/Hz}). This performance is on par with the highest-linearity silicon-based modulators previously reported (\qty{123}{dBHz^{2/3}}) \cite{Yu2020HighDistortion}, as well as the bulky \ce{LiNO_3} MZM (\qty{123}{dBHz^{2/3}}) \cite{Daulay2022UltrahighFilter}, while maintaining a much simpler system.

\subsection{SiGe EAM modulator}

\begin{figure}
    \centering
    \includegraphics[width = \linewidth]{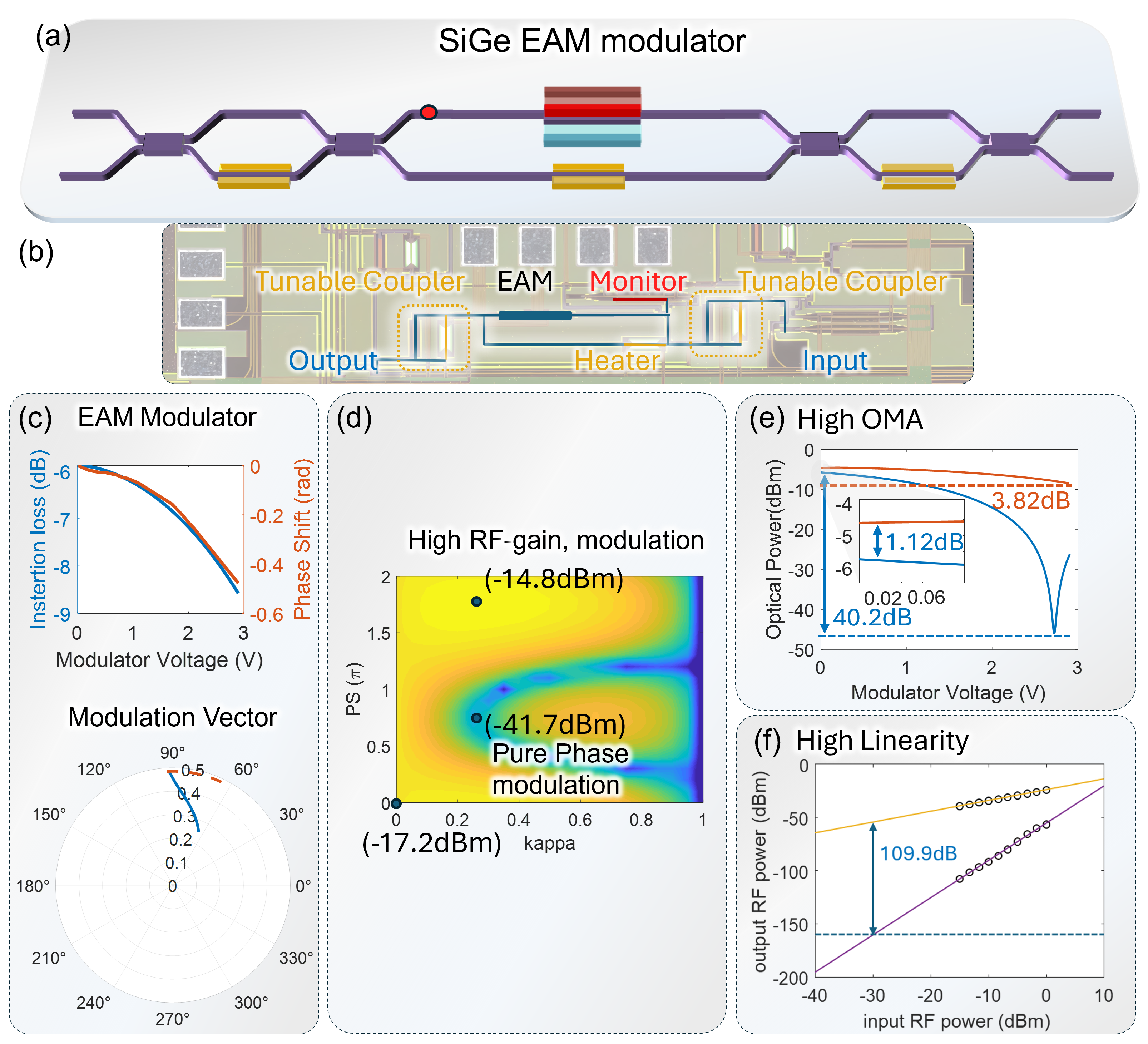}
    \caption{Programmable modulator circuit based on SiGe EAM modulator. (a) The schematic of the modulator; (b) Fabricated demonstrator; (c) Insertion loss and the phase shift introduced by the embedded SiGe modulator;(d) Measured modulated signal power with a sweep of kappa and offset phase shift;  (e)DC response of the optimized high extinction modulation and a normal SiGe modulator; (f) Optimized high-linearity modulation.}
    \label{f:3}
\end{figure}

In silicon photonics, a \ce{SiGe} EAM is frequently utilized as an intensity modulator due to its compact size and broad modulation bandwidth. However, it is hindered by significant insertion loss and low modulation efficiency. By integrating it into our programmable modulator circuit, we can enhance the performance of this SiGe EAM. The schematic of the modulator circuit is shown in Fig.~\ref{f:3}(a), and a fabricated demonstrator is shown in Fig.~\ref{f:3}(b). The characterized DC response of an independent \ce{SiGe} EAM is illustrated in Fig.~\ref{f:3}(c). 

\paragraph{High RF-gain, low chirp intensity modulation}
The generated RF gain depends on the modulation slope and the optical insertion loss. In our programmable modulator circuit, if the light is partially fed into the other arm (not the EAM arm), the overall insertion loss of the system will be lower. Meanwhile, the intensity modulation efficiency itself decreases, as less light is fed into the EAM. However, the spurious phase modulation (although it is very weak) can also be converted into intensity modulation, which will also lead to a lower chirp. As shown in Fig.~\ref{f:3}(d), when the programmable modulator is set as  $\kappa = 0.3$ and $\phi_s = 1.9\pi$, the generated RF signal can be around \qty{2.6}{dB} higher than the EAM by itself (when biasing at \qty{0.5}{V}). Meanwhile, the insertion loss of the programmable modulator is \qty{0.6}{dB} lower than that of the single SiGe EAM, which also contributes to the improved RF gain.

\paragraph{High extinction intensity modulation}
SiGe EAMs also suffer from low modulation extinction. Taking advantage of the phase modulation of the EAM, our programmable modulator can be set to have a high-extinction-ratio modulation. Experimental results in Fig.~\ref{f:3}(e) show that the optical intensity variance can reach \qty{40}{dB} extinction ratio within the applied voltage range, which is \qty{36}{dB} higher than a single SiGe EAM (\qty{3.82}{dB} at \qty{1550}{nm}), with an \qty{1.12}{dB} extra insertion loss (at \qty{0}{V}) as a trade-off.

\paragraph{High-linearity intensity modulation}
As shown in Fig.~\ref{f:3}(b), the intensity response is also not linear. Our programmable modulator circuits can also improve linearity. Experimental results in Fig.~\ref{f:3}(f) show that the SFDR of the system can be boosted to \qty{110}{dBHz^{2/3}} ($\kappa = 0.55, \phi_s = 0.9\pi$), which is around \qty{22}{dB} higher than the standalone SiGe EAM (\qty{88}{dBHz^{2/3}}, biasing at \qty{0.5}{V}). In other tests, the EAM SFDR can be improved from 102.8 to \qty{112.5}{dBHz^{2/3}} when biasing at \qty{1}{V}, or 109 improved to \qty{114.5}{dBHz^{2/3}} when biasing at \qty{2}{V}. Compared with previously reported results \cite{Hraimel2011ExperimentalLinks,VanGasse2019SiliconUp-conversion}, our programmable modulator shows higher SFDR and better linearity optimization. More details are listed in the Supplementary information.

\subsection{Microring resonator modulator}

Previous experimental results show the optimization possibilities of the proposed programmable modulators. Another commonly used optical modulator in the silicon photonics platform is the microring modulator (MRM), known for its compact size, low power consumption, and high operational bandwidth. However, it can only work close to its resonance wavelength, and its intensity modulation is quite nonlinear. Here, we extract the DC responses of an MRM fabricated in the imec ISIPP50G platform, develop a quasi-static model of the MRM, and simulate its RF responses when embedded within our programmable modulator. At the time of writing, we have not implemented experimental circuits for the MRM.

\begin{figure}
    \centering
    \includegraphics[width = \linewidth]{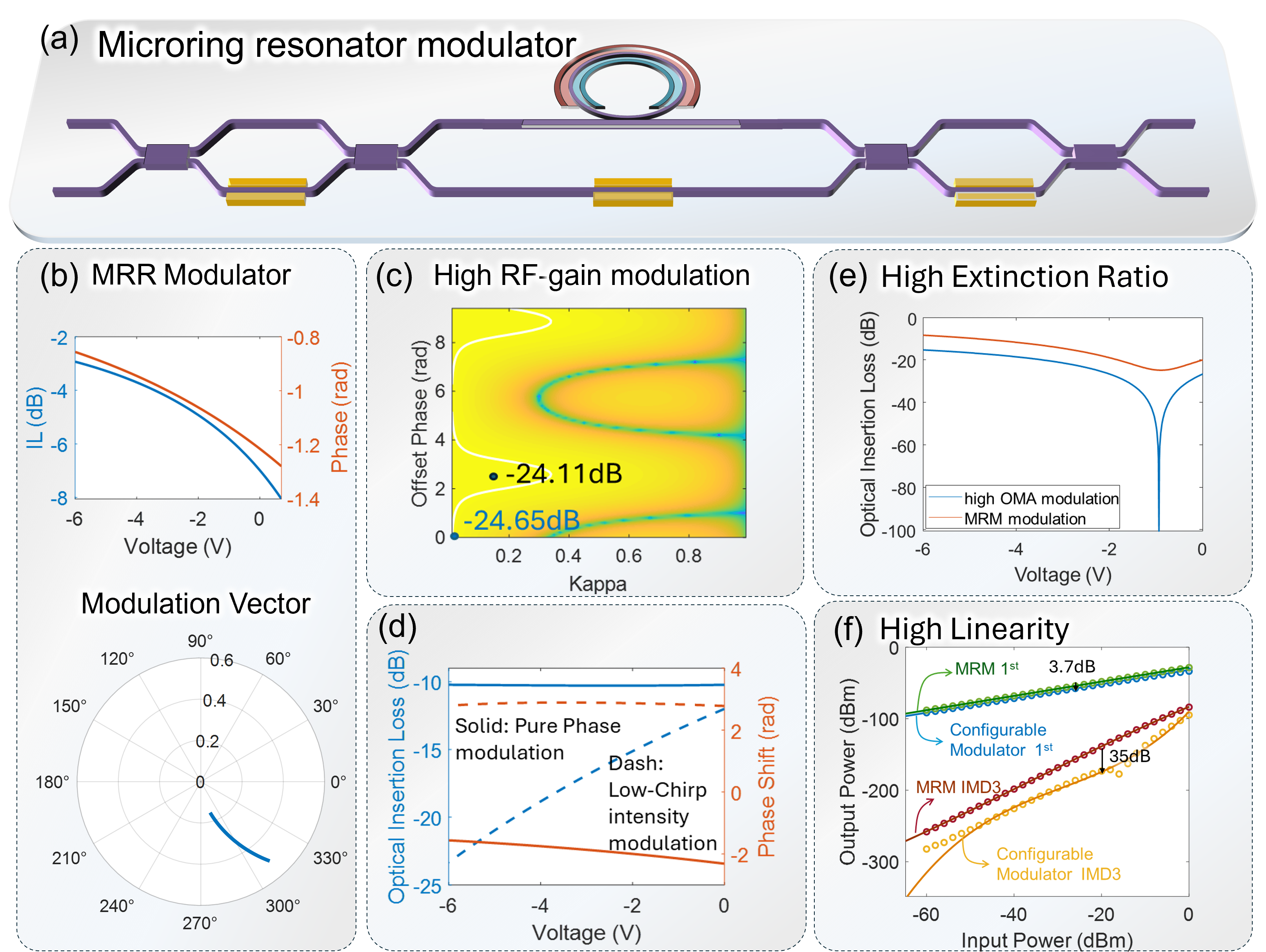}
    \caption{programmable modulator based on a microring modulator. (a) The schematic of the modulator; (b) Insertion loss and the phase shift introduced by the embedded microring modulator. }
    \label{f:4}
\end{figure}

As a cavity, an MRM exhibits different responses at varying wavelengths \cite{Bogaerts2012SiliconResonators}. The fabricated MRM was measured with a Q-factor of 2212.1 and a resonance wavelength of \qty{1556.424}{nm}. For the simulation, we set the operating wavelength to \qty{1556.2488}{nm}, where the MRM generates the strongest RF signal. The measured MRM modulation responses of the MRM at this wavelength are shown in Fig.~\ref{f:4}(b).

\paragraph{High RF-gain intensity modulation}

To simulate the RF response using our MRM model, we applied a sinusoidal waveform with an amplitude of \qty{1}{V} and a bias voltage of \qty{-1}{V} as the input RF signal for the MRM. By sweeping the optical wavelength, we identified  \qty{1556.2488}{nm} and \qty{1556.655}{nm}, which yielded the highest optical sidebands. These wavelengths are located on opposite sides of the resonant wavelength. At these points, the optical phase variation exhibits opposite behavior: one phase increases with a higher reverse bias voltage, while the other decreases with the same voltage increase. Here we used \qty{1556.2488}{nm} as the carrier wavelength for these simulation analysis.

To further improve the RF-gain, we can embed the MRM in the proposed programmable modulator. By setting $\kappa = 0.13$ and $\phi_s = 0.828\pi$, we can pump up the generated RF signal by another \qty{0.5}{dB}, as shown in Fig.~\ref{f:4}(c).

\paragraph{Low-Chirp intensity modulation}

As shown in Fig.~\ref{f:4}(b), intensity modulation from an MRM would also come with phase modulation, which would also introduce frequency chirping in the modulated light signal. Here again, we can set our programmable modulator with kappa and offset phase, thus the phase change can be reduced from from \qty{0.3585}{rad} to \qty{0.1109}{rad} in the bias range of \qty{6}{V}, and the intensity variance can also be improved from \qty{4.0547}{dB} to \qty{11.3301}{dB}, while trade off a \qty{5}{dB} higher insertion loss when the MRM is biased at \qty{0}{V}, as shown in Fig.~\ref{f:4}(d).

\paragraph{Pure phase modulation}
Oppositely, we can also set the programmable modulator with $\kappa = 0.32$ and $\phi_s = 1.94\pi$. At this operation point, the intensity variance is reduced from \qty{4.0547}{dB} to \qty{0.0664}{dB} in the reversed voltage range of \qty{6}{V}, while the phase variance can even be improved from \qty{0.3585}{rad} to \qty{0.7545}{rad}, as shown in Fig.~\ref{f:4}(d).

\paragraph{High extinction intensity modulation}
As an optical cavity, the designed MRM can already provide a high extinction intensity modulation. The measured extinction ratio of the fabricated MRM is around \qty{26}{dB}. While, in the simulation, we can precisely set the programmable modulator ($\kappa = 0.3372$ and $\phi_s = 1.638\pi$) to have an infinite extinction, where the embedded MRM can only provide \qty{16.5}{dB} (optical carrier is at \qty{1556.454}{nm} at the moment), as shown in Fig.~\ref{f:4}(e). But this operation would introduce around \qty{7}{dB} extra optical insertion loss as a trade-off.

\paragraph{High-linearity intensity modulation}

The linearity of the MRM highly depends on the optical signal wavelength for the modulation, and it is also quite limited, due to its Lorentzian filtering response. As discussed above, we set the optical wavelength at \qty{1556.2488}{nm} to ensure the highest RF gain. From the simulation, we can find out that, when programmable modulator is set at $\kappa = 0.35$ and $\phi_s = 0.414\pi$, the SFDR can be improved around \qty{10}{dB}, with an \qty{3.7}{dB} lower RF gain, as shown in Fig.~\ref{f:4}(h).

\section{Discussion}\label{sec4}

Our programmable modulator circuit embeds the high-speed modulator into a tunable MZI structure, and uses the static MZI response as well as the high-speed phase response (intensity response) to optimize the intensity response (phase response) of the embedded modulator. As demonstrated earlier, the enhanced performances are exceptional, as listed in Table~\ref{t:1}, although there are still some constraints and opportunities for further improvement.

\begin{table}[]
\caption{Proposed modulator circuit performance} \label{t:1} 
\begin{tabular}{|l|l|l|l|l|l|}
\hline
\rowcolor[HTML]{68CBD0} 
{\color[HTML]{FFFFFF} \textbf{Modulator Circuit}} &
  {\color[HTML]{FFFFFF} \textbf{}} &
  {\color[HTML]{FFFFFF} \textbf{\begin{tabular}[c]{@{}l@{}}Spurious IM \\ (dB) ***\end{tabular}}} &
  {\color[HTML]{FFFFFF} \textbf{\begin{tabular}[c]{@{}l@{}}RF Gain \\ (dB)\end{tabular}}} &
  {\color[HTML]{FFFFFF} \textbf{\begin{tabular}[c]{@{}l@{}}Extinction \\ (dB)\end{tabular}}} &
  {\color[HTML]{FFFFFF} \textbf{\begin{tabular}[c]{@{}l@{}}SFDR \\ (dBHz$^{2/3}$)\end{tabular}}} \\ \hline
\rowcolor[HTML]{C0C0C0} 
\cellcolor[HTML]{C0C0C0} &
  {\color[HTML]{000000} Performance*} &
  {\color[HTML]{000000} \textless{}-70} &
  {\color[HTML]{000000} -18} &
  {\color[HTML]{000000} 38} &
  {\color[HTML]{000000} 124.6} \\ 
\rowcolor[HTML]{EFEFEF} 
\multirow{-2}{*}{\cellcolor[HTML]{C0C0C0}PN depletion based} &
  Improved** &
  -23 &
  0.7 &
  21.5 &
  15 dB \\ \hline
\rowcolor[HTML]{C0C0C0} 
\cellcolor[HTML]{C0C0C0} &
  Performance* &
  \textless{}-41.7 &
  -14.8 &
  40.2 &
  109.9 \\ 
\rowcolor[HTML]{EFEFEF} 
\multirow{-2}{*}{\cellcolor[HTML]{C0C0C0}SiGe EAM based} &
  Improved** &
  -- &
  2.4 &
  36.38 &
  22 dB \\ \hline
\rowcolor[HTML]{EFEFEF} 
\cellcolor[HTML]{C0C0C0}MRM based (simulation) &
  Improved** &
  -40 &
  0.5 &
  80 &
  10 dB \\ \hhline{|=|=|=|=|=|=|}
\rowcolor[HTML]{DFDFDF} 
Bulky LN Modulator \cite{Daulay2022UltrahighFilter} &
  Performance & \textgreater{}-64 
   & 1.2
   & \textgreater{}20
   & 123
   \\ \hline
\rowcolor[HTML]{DFDFDF} 
Microring Modulator \cite{Cheng2023On-chipSensing}  &
  Performance & --
   & --
   & 71
   & --
   \\ \hline
\rowcolor[HTML]{DFDFDF} 
PN-Junction \cite{Bottenfield2020High-PerformanceSubsystems}  &
  Performance & --
   & -19.5
   & 45
   & 92
   \\ \hline
\rowcolor[HTML]{DFDFDF} 
SiGe EAM \cite{VanGasse2019SiliconUp-conversion}  &
  Performance & --
   & -20
   & 45
   & 82
   \\ \hline
\end{tabular}
\begin{tablenotes}
\footnotesize
\item[* The optimized configuration is different for the metrics.]
\item[** Compared with the embedded modulator.] 
\item [*** Lower is better.]

\end{tablenotes}
\end{table}

\begin{figure}
    \centering
    \includegraphics[width = \linewidth]{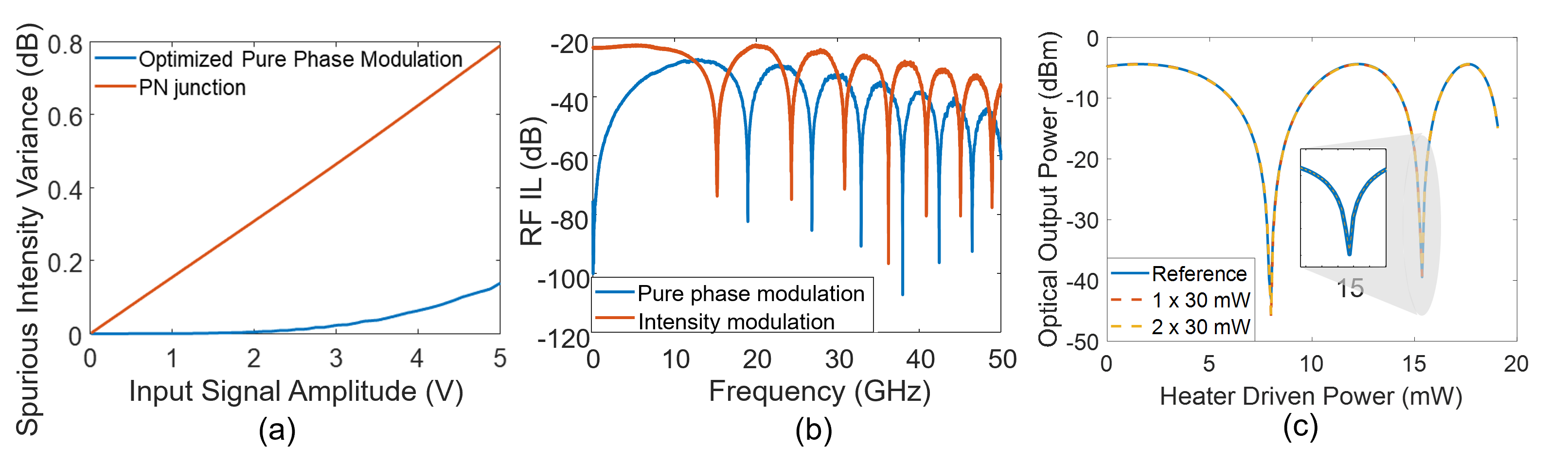}
    \caption{(a)Large signal analysis, with PN junction based programmable modulator for pure phase modulation; (b)Pure phase modulation verification, two microwave photonic filters with programmable modulator and \qty{25}{km} fibers. (c) Thermal crosstalk analysis. }
    \label{f:5}
\end{figure}

\paragraph{Large signal performance} 
Due to the low modulation efficiency of silicon photonic modulators, they typically require large RF drive signals. This requirement may limit the optimization performance of the proposed programmable modulator. Here, we revisit the PN junction-based programmable modulator as an example. As previously discussed, its spurious intensity modulation can be effectively suppressed. Fig.~\ref{f:5}(a) presents the simulated intensity variance as a function of input power for the programmable modulator with a PN junction and the embedded PN junction. The results demonstrate that while the optimization effect diminishes with increasing input signal power, it still provides a noticeable performance enhancement.

\paragraph{Pure phase modulation verification}
We demonstrated that, under specific configurations, our programmable modulator is capable of delivering pure phase modulation with minimal intensity variation. However, direct detection of the light's phase variance is not feasible using standard photodiodes. To address this, we employed a \qty{25}{km} optical fiber, leveraging its dispersion to create a two-tap microwave photonic filter \cite{Yao2009MicrowavePhotonics}. The filtering results are presented in Fig.~\ref{f:5}(b). The two filter responses are interleaved, revealing that the light signals exhibit phase modulation and intensity modulation, respectively. This confirms the ability of our programmable modulator to operate as a broadband RF filter with a tunable extinction ratio, achieving a maximum of \qty{80}{dB}.

\paragraph{Optical length matching}
Another factor that can influence the optimization behavior is the optical length balance in the proposed programmable modulator. However, perfect balance between the two arms of the modulator is not achievable due to the asymmetric optical design, where one arm contains the modulator while the other does not. We designed our circuits in a symmetric way, but it still shows some dispersion introduced by the optical delay difference. This dispersion would add extra phase difference in the optical modulated sidebands, distorting the setting points needed of the programmable modulator circuits. And this is also the reason why, in Fig.~\ref{f:5}(b), the maximum points of the intensity modulation cannot fully match the minimun points of the phase modulation.

\paragraph{Thermal crosstalk}
Our demonstrators for the modulators configurations are based on the tuning of our thermal optical phase shifts, which needs extra driven electrical powers and introduced thermal crosstalks (main contributing to the phase mismatch between simulation and experimental results). To reduce the crosstalk, the thermal phase shifters we used in the demonstrators are enhanced with a local substrate undercut, which help increasing the thermal efficiency and reducing the thermal crosstalk simultaneously. We tested the thermal efficiency and crosstalk of the heaters. First, we swept the driving power applied to a phase shifter in a tunable coupler and monitored the optical output. Next, we activated a neighboring phase shifter with \qty{30}{mW} of power and repeated the sweep, followed by activating an additional neighboring phase shifter and performing the sweep once more. The resulting optical power traces are plotted in Fig.~\ref{f:5}(c). As shown, the phase shifter demonstrates a thermal efficiency of \qty{3.68}{mW/\pi}, and the crosstalk effects on the coupling ratio are negligible.

An alternative way is that the circuits can be designed and fabricated statically with the specific parameters that match the desired optimized configuration. Like in \cite{Proesel2019Nanosecond-scaleSwitch}, a switch was proposed for no residual amplitude modulation, and it was specifically designed and fabricated with optimized parameters which are discussed in Section 2.1 in this paper. Additionally, innovative approaches such as phase-changing materials, MEMS, or liquid crystal materials might be employed here to achieve low power consumption and minimal thermal crosstalk in optical phase tuning.

\section{Conclusion}\label{sec5}

In this paper, we present a programmable modulator circuit design that achieves optimized performance for the embedded modulator device. The optimization is carried out by configuring the static phase shifters within the circuits, with the embedded modulator being the sole high-speed component. Consequently, the high-speed performance (\qty{3}{dB} bandwidth) remains unaffected by the configuration. Unlike other optimization methods, our programmable modulator circuit structure is simple and straightforward, ensuring that simulation results closely match experimental outcomes. Furthermore, our programmable modulator design is compatible with all integrated photonic platforms, making it a universal optimization method for all modulator devices. In this paper, we made two real demonstrations with traveling wave \textit{pn} junction and \ce{SiGe} EAM modulators, and showed the simulation results with MRR in silicon photonic platform. The optimized programmable modulator circuits with \textit{pn} junction can offer the best pure phase modulation, as far as we know, and the optimized linearity of two demonstrators also reach the best results ever presented. Therefore, this modulator design has the potential for widespread use in optical communication, LiDAR, microwave photonics, and other systems, addressing the growing demands of various applications.

\bmhead{Supplementary information}

See the attachment file.

\bmhead{Acknowledgments}

The research presented here was supported by the European Research Council through the Consolidator Grant PhotonicSWARM (grant 725555), and the European Horizon2020 program through the projects MORPHIC (grant 780283) 

\bmhead{Author Contributions} H.D., XF.C. and W.B. conceived the idea of this work. H.D. simulated, designed and laid out the whole system.  H.D. and Y.Z. performed the experimental characterizations and analysis. H.D., Y.Z. and W.B. wrote the paper.

\section*{Declarations}
The authors declare no conflicts of interest

\bibliography{references.bib}% common bib file

\section{Methods}\label{sec3}

\paragraph{Silicon chip fabrication.} The measured modulators was fabricated in the imec iSiPP50G silicon photonic platform on \qty{200}{mm} wafers, in which \qty{30}{GHz} PN junction based modulators, \qty{50}{GHz} MRM, and \qty{50}{GHz} SiGe EAM is provided. The chip layout is designed using IPKISS by Luceda Photonics. 

\paragraph{System driving and characterization}

In our demonstrations, the thermal phase shifters, the tap monitors, and the modulator DC bias, are driven by source meters (Keithley 2401). The optical carrier is provided by a tunable laser source (Santec TSL550), and the recovered RF signal is generated by a \qty{42}{GHz} PD (Discovery LabBuddy DSC10H). The RF response of the optical link is measured by a vector network analyzer(Keysight E8364B, \qty{50}{GHz}). The RF signal sources for the SFDR tests are two signal generators (Rohde\&Schwarz SMR 40) and the used electrical spectrum analyzer is a Keysight EXA signal analyzer (N9010A \qty{44}{GHz}).

\paragraph{System calibration}
Optical phase shifters play an important role in our programmable modulator design. The imec iSiPP50G silicon photonic platform provides doped silicon heater with undercut process, enjoying high thermal efficiency and low thermal crosstalk. The measured thermal efficiency $P_{\pi}$ is around \qty{9}{mW}. In our demonstrators, we introduced tapped photodetectors to monitor the optical power after every tunable couplers. By tuning every heater, and checking the output power of the tunable coupler, we can calibrate the whole system, and build up a look-up table for setting the system.

\paragraph{Embedded Modulator characterization}
The performance of the programmable modulator is determined by the embedded E/O modulator, as well as the splitting ratio of the tunable couplers (which can be tuned from 0 to 1) and the offset phase shift (which can be set between 0 to $2\pi$). Thus, the characterization of the embedded E/O modulator is critical.

The embedded modulator can be directly characterized using the proposed programmable modulator. The reference performance can be obtained by setting the splitting ratio of the tunable couplers to 1, ensuring no light is fed into the embedded modulator. Conversely, the intensity response of the embedded modulator can be measured by setting the tunable couplers to 0, directing all light into the embedded modulator.  We can then configure the tunable couplers to a 50:50 ratio to form a Mach-Zehnder interferometer and use the interference behavior, along with the previously extracted intensity response, to fit the phase response of the embedded modulator. This method indicates that our programmable modulator can also be designed to characterize an unknown optical components.

\section*{Data availability}
.

\section*{Code availability}

\end{document}